# A label-free method to quantify early-stage amyloid aggregation *via* intrinsic phenylalanine fluorescence

Gaëlle Audéoud[1,2,§], Louis Moine[2,§], Laura Bonnecaze[2], Maxime Lavaud[1,2,3], Lucie Khemtemourian[2], Yacine Amarouchene[1], Thomas Salez[1,*], Marion Mathelié-Guinlet[2,*]

[1] Univ. Bordeaux, CNRS, LOMA, 33400 Talence, France

[2] Univ. Bordeaux, CNRS, Bordeaux INP, CBMN, UMR 5248, F-33600 Pessac, France

[3] Faculty of Physics, University of Vienna, Boltzmanngasse 5, 1090 Vienna, Austria

[§] These authors contributed equally

[*] Corresponding authors: marion.mathelie-guinlet@u-bordeaux.fr; thomas.salez@cnrs.fr

**Abstract.** The aggregation of amyloid-forming peptides is a dynamic, complex process that underlies their diverse biological activities, from physiological functions to disease-associated dysfunctions. While the structure of fibrillar end-products is well-characterized for most amyloids, the heterogeneous and often transient oligomers, likely key in cytotoxicity, remain poorly investigated, especially for peptides with low-yield aromatic residues. Here, by exploiting and developing flow induced dispersion analysis in both peak and front modes, we demonstrate that intrinsic phenylalanine fluorescence can be harnessed to quantify the conversion of diffusing monomers into non-diffusing oligomers and fibrils. This approach is validated using PSMα3, a tryptophan- and tyrosine-free peptide from *S. aureus*, known for its fast amyloid fibrillation and key roles in bacterial virulence. We indeed characterize its low-molecular-weight oligomers, and size evolution from 2 to 10 nm over time. Importantly, to further validate the robustness of our front-mode approach, we apply it to hIAPP, a tyrosine-containing, tryptophan-free peptide, with critical outcome in type 2 diabetes. Our results overcome the limitations of traditional biochemical and biophysical amyloid assays by extending analysis from large oligomers and fibrils to small heterogeneous oligomers, under near-physiological conditions. This study thus offers a new analytical framework, thereby filling a critical gap in amyloid research, to probe the early stages of aggregation, key in the design of alternative therapeutics for amyloid-diseases.

## Introduction

Amyloid-forming peptides and proteins are ubiquitous in nature, present in several living organisms from humans to bacteria, and are characterized by their ability to aggregate from monomeric units into oligomeric intermediates and *in fine* highly ordered fibrils. They are not only the hallmark of more than 50 human diseases[1–3], including neurodegenerative disorders such as Alzheimer and Parkinson's diseases[4], but have also been identified as functional - beneficial - components for biological systems[5,6]. Particularly in bacteria, functional amyloids play key roles in virulence and infection mechanisms[7–9]. Despite sequence and function diversity, all amyloids share a common structural scaffold, the so-called cross-β architecture, resulting from the self-assembly of mono- and oligomeric subunits *via* backbone hydrogen bonding and side-chain interactions[10,11].

Amyloid fibrillation is a highly dynamic and environment-dependent process with dynamic exchange between initial monomers, intermediate oligomers and end-products - the fibrils - each displaying particular properties associated to different biological activities. While fibrils were initially thought to account for amyloids cytotoxicity[12,13], increasing evidence suggests that oligomeric species, formed during the early stages of aggregation, may instead be the membrane-active species responsible for toxicity[14–18]. However, the highly dynamic, heterogeneous, and eventually transient, nature of populations co-existing along the pathway, akin to the metastable fibre states arising from geometrical frustration in self-assembling model systems[19], represent a major analytical challenge, as their quantitative measure remains elusive with conventional techniques. While cryo-Electron Microscopy (cryo-EM) and solid-state Nuclear Magnetic Resonance (ssNMR) have granted breakthroughs in the structural characterization of amyloid fibrils at the atomistic level [20,21], both struggle to capture the transient oligomeric species, and often require non-physiological preparations. On the other hand, Dynamic Light Scattering (DLS) and size-exclusion chromatography can probe hydrodynamic sizes of distinct species in solution, yet are prone to artefacts from polydisperse and heterogeneous samples. Likewise, Atomic Force Microscopy (AFM) achieves nanometer-resolution imaging of those species but only on surface-bound samples[22,23]. Finally, fluorescence-based methods, such as Thioflavin T (ThT) fluorescence spectroscopy[24,25], or fluorescence microscopy[26–28], remain central for kinetic studies of amyloids, but essentially lack the spatial resolution to probe the early stages of aggregation, and often require extrinsic labelling, potentially perturbing the process. Alternatively, they rely on the intrinsic fluorescence of tryptophan (Trp) or tyrosine (Tyr) residues (excitation and emission wavelengths at $\lambda_{abs}$~ 280 nm, $\lambda_{em}$ ~ 350 and 300 nm resp.; quantum yield ~ 0.2)[29,30], which might be lacking in proteins of interest. Consequently, despite the complementary strengths of these methods, there is still a strong need for label-free, solution-phase analytical methods capable of tracking early amyloid species - likely responsible for toxicity and thus diseases - under minimally perturbing conditions.

In this context, Flow-Induced Dispersion Analysis (FIDA) emerges as a promising microfluidic technique for the quantitative determination of hydrodynamic size and separation of distinct species populations within heterogeneous samples[31]. Briefly, it is based on Taylor dispersion[32,33], *i.e.* the enhanced physical diffusion of solutes under laminar shear flow, from which one can derive molecular diffusion coefficients, and in turn hydrodynamic radii, without size bias as solutes contribute to the signal proportionally to their mass abundance[34] (Fig. 1). Interestingly, this technique requires very limited sample volumes (<1 µL), no pretreatment, and allows real-time monitoring of objects ranging from angström to sub-micron scales. Recent studies have highlighted the potential of FIDA in quantifying monomeric, and oligomeric species during aggregation of Aβ[35,36] and α-synuclein[37], as well as in liquid-liquid phase separation (LLPS) characterization[38]. Nonetheless, while FIDA has been successfully, yet scarcely, applied to amyloid proteins, its use has so far been restricted to intrinsic fluorescence of Trp or Tyr or extrinsic labelling, with development such as FibrilPaint for Tau fibrillation[39]. However, such a labelling may alter the aggregation behaviour and kinetics, especially in the case of small peptides[40]. This limitation is understandable as phenylalanine (Phe), although ubiquitous in amyloid proteins, exhibits a very low fluorescence quantum yield (~ 0.04) – and emits fluorescence two to three orders of magnitude lower than Trp or Tyr[29,30]. It is thus generally considered unsuitable for analytical detection.

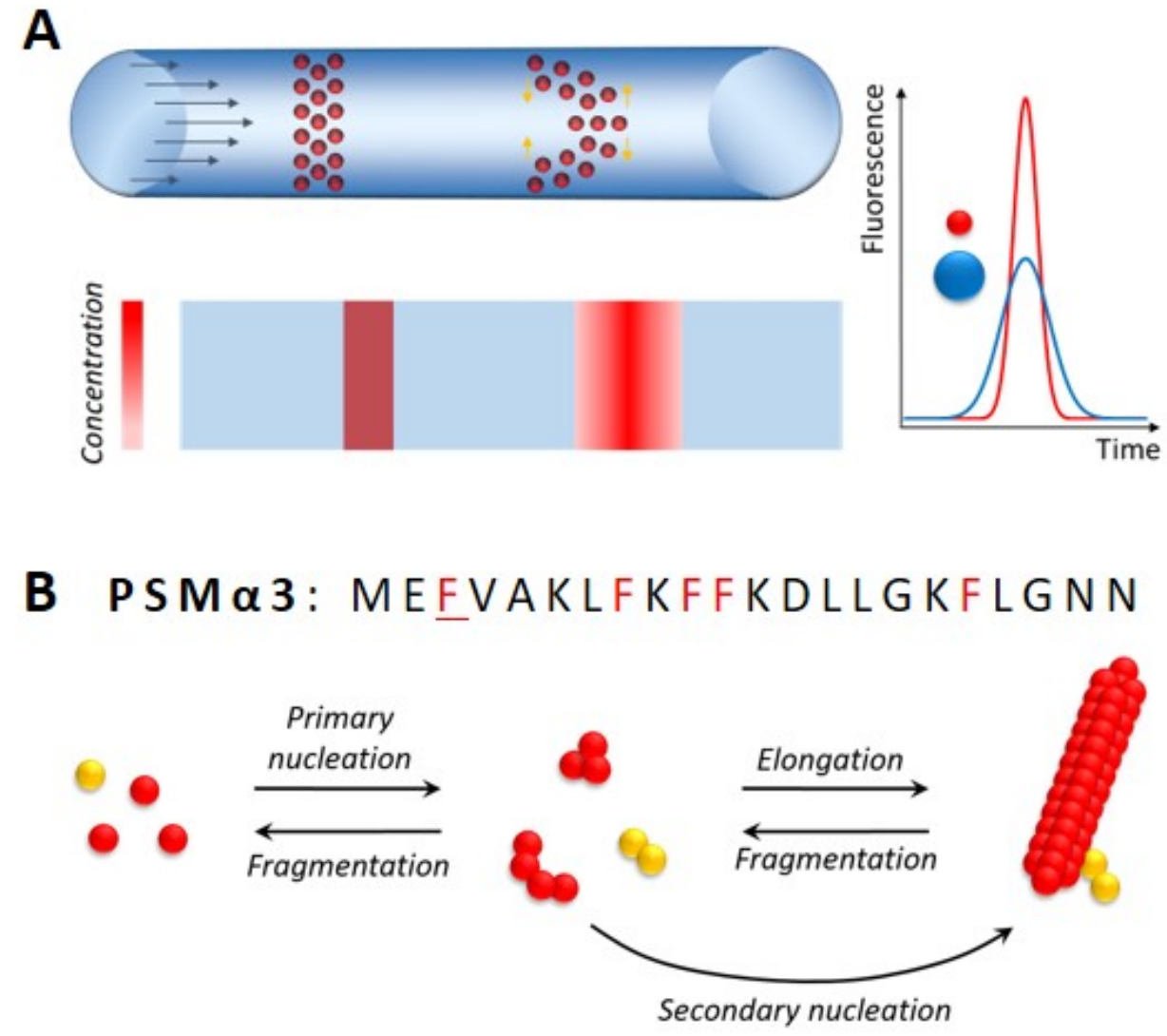

**Figure 1. Flow-induced Dispersion Analysis for amyloid speciation. (A)** Schematic of the Taylor dispersion analysis principle. When injected in a microfluidic channel, a suspension of particles is advected by the parabolic Poiseuille flow, leading to an enhanced longitudinal dispersion according to their size. In peak mode, a plug of particles is injected, and detected by a fluorescence detector (excitation wavelength, λ = 280 nm) after a given delay in time, $t_0$. Under large Peclet regime, large particles are more prone to dispersion, leading to broader elution peak, called taylorgram, than small particles. **(B)** Amino-acid sequence of the studied peptide, the wild-type (WT) PSMα3, where phenylalanine residues are highlighted in red and the mutation site (F3A) is underlined. A theoretical representation of amyloid aggregation is also shown.

In this work, we demonstrate that intrinsic Phe fluorescence can be harnessed, after appropriate optimisation, to monitor the *in vitro* aggregation process of amyloid proteins, using FIDA. As a model system, we selected the phenol soluble modulin alpha 3 (PSMα3), a 22 amino acid-long peptide secreted by pathogenic *Staphylococcus aureus*[41] (Fig. 1B). Beyond its medical relevance, as a key player in biofilm formation, immune evasion and cytotoxicity[42,43], PSMα3 is an ideal model since its fibrillation dynamics has been so far overlooked due to its exceptionally fast kinetics (< 2 hour)[44,45] and its lack of Trp and Tyr residues. To address these challenges, we developed a front-mode approach, beyond classical peak-mode FIDA, coupled with an algorithmic analysis, as a novel quantitative tool for comprehensive picture of amyloid speciation in near-physiological conditions. This approach allowed us to quantify a two-phase consumption of PSMα3 species within 2 hours, accompanied by an increase in hydrodynamic radii from ~2 to 4 nm, and an increasingly larger proportion of insoluble oligomers and fibrils. Our approach thus provides a novel analytical framework to characterize fast-assembling, Trp/Tyr-free proteins, especially the kinetic and morphological features of the species formed during the process. Given the established importance of such species in well-studied proteins, and their putative roles in less investigated peptides, such a characterization could help decipher why these species, rather than mature fibrils, are involved in amyloid toxicity mechanisms.

## Materials and methods

**Materials.** Formylated PSMα3 peptides were purchased from GenScript at a purity > 98 %. Human IAPP (hIAPP) was purchased from Bachem at a purity > 95%. Trifluoroacetic acid (TFA, Z99% HPLC grade) was purchased from Fisher Scientific, and thioflavin T (ThT) and hexafluoroisopropanol (HFIP) from Sigma Aldrich.

**Peptide preparation.** PSMα3 peptides (powder) were first pre-treated in a (1:1) mixture of HFIP/TFA, at a final concentration of 1 mmol/L for one hour. The solvent was then evaporated under a stream of dry $N_2$, and further let under vacuum in a desiccator for at least two hours. The obtained peptide film was either stored at - 80 °C, or directly used for experiments by rehydration at the desired concentration in the appropriate medium – unless otherwise stated, 20 mmol/L sodium phosphate buffer containing 100 mmol/L NaCl, pH 7.4. Preparation of hIAPP followed the same steps, with a pretreatment in HFIP alone.

**Flow Induced Dispersion Analysis.** FIDA experiments were performed on a Fida Neo platform, that uses a LED excitation as the light source, to detect the fluorescent signal from our label-free peptides, which emit fluorescence from phenlylanine residues. The excitation light (280 nm) is coming from the top, and the emission light - from 320 to 400 nm – is collected in the same path, and amplified through photomultiplier tubes (PMT). We used UV-transparent fused silica capillaries of 75 µm in diameter and 100 cm in length (Fidabio, https://escooptics.com/pages/materials-fused-silica-quartz). Those capillaries were permanently coated on the inner wall, to reduce peptide sticking; while the coating composition is propriety, it is UV-transparent too. Samples were loaded in a 96-well plate, placed in an autosampler which directly injects the appropriate sample on the inlet end of the capillary, at a given injection pressure and following a Poiseuille parabolic flow. The sample then, upon the combination of both flow and molecular diffusion, pass through a UV detection window at 80 cm from the inlet end. Two modes were used in this study:

- the peak mode: after equilibration with sample buffer (20 mmol/L sodium phosphate buffer with 500 mmol/L NaCl), a sample plug was injected at 50 mbar for 10 seconds. It was then pushed through the capillary by the buffer at a given mobilizing pressure between 400 and 800 mbar (except for calibrations).

- the front mode: after equilibration with sample buffer, the sample is continuously loaded in the capillary at a given mobilizing pressure between 400 and 800 mbar (except for calibrations) for approx. 5 minutes.

Importantly, the mobilizing pressure was chosen depending on the peptide of interest. While large species require more time to diffuse and therefore a low mobilizing pressure, small species would diffuse faster and would be detected even at high pressure (Fig. 2) following this optimization condition[51]:

$$R_h \leq \frac{0.34 k_B T l L}{\Delta P R_c^2}$$

Consequently, a high pressure of 800 - even 1600 mbar - was selected for (i) calibration studies where peptides are not screened in time and for (ii) the mutant F3A peptide that does not aggregate in time, remaining as monomers of ~1.3 nm in theoretical hydrodynamic radius. Conversely, for the WT peptide known for its fast fibrillation, a pressure of 400 mbar was chosen, seen as a good compromise to probe particles of up to 14 nm in less than five minutes. During this time window we hypothesize that the peptide will not undergo aggregation in the capillary.

In both modes, the capillary was initially flushed with 1 mol/L phosphoric acid followed by a flush with ultrapure water. Then, between each measurement, the capillary was rinsed with a 20 mmol/L sodium phosphate buffer with 500 mmol/L NaCl. Experiments were performed either (i) in real-time on the same sample loaded in a single well, (ii) or on pre-incubated samples disposed in distinct wells and measured for a given time, at room temperature. In the latter case, samples were prepared on bench and fast-frozen or directly measured at given time points.

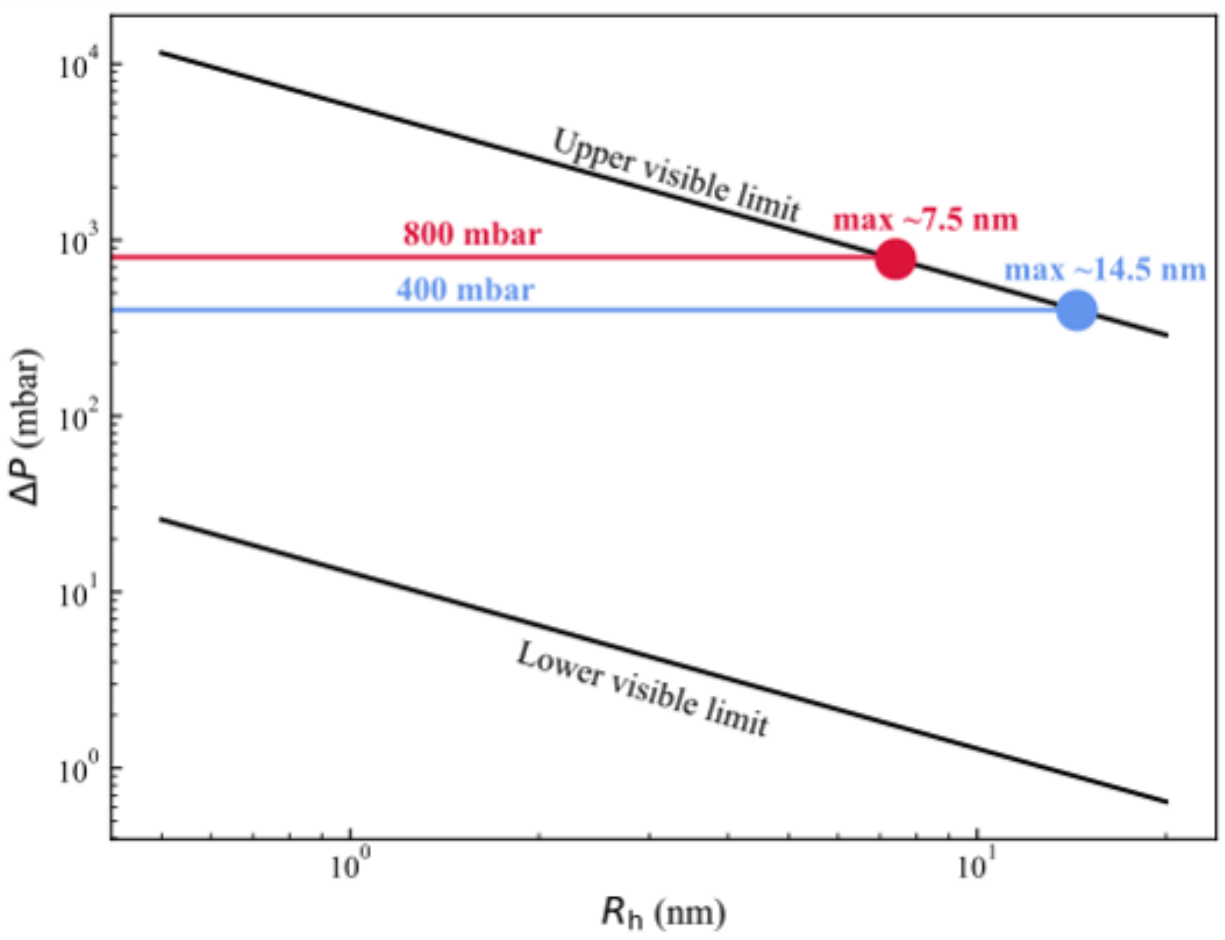

**Figure 2. Optimal mobilizing pressures for Flow-induced Dispersion Analysis.** Theoretical limits of detection for particles of hydrodynamic radius $R_h$ as a function of the mobilizing pressure. The capillary length is 100 cm, and its diameter 75 µm.

**Taylorgram analysis.** As detailed previously, molecular diffusion coefficients, and thereby hydrodynamic radii, can be obtained by appropriately fitting the FIDA elution profiles, or taylorgrams, resulting from the combination of the Poiseuille flow and the diffusion of all species in the capillary. Of note, while in peak mode, the taylorgram has a typical Gaussian shape, in front mode it appears as a sigmoidal-shaped curve (see details above). For the former, the fit was applied to ~50% of the curve to account for the asymmetric shape resulting from peptide sticking to the capillary wall. For the later, between 70% and 90% of the curve was fitted, while we verified that the percentage chosen did not change the derived sizes. If one considers a polydisperse solution where distinct size populations co-exist, as it is the case for an aggregating-peptide, a taylorgram cannot result from a single hydrodynamic radius probability density function, $P(R_h)$, rendering the fitting procedure more complex. We thus used the Constrained Regularized Linear Inversion (CRLI) approach of Cipelletti *et al.*[52], and inspired by the CONTIN algorithm[53,54] widely used in Dynamic Light Scattering (DLS). Briefly, this method imposes the size distribution to be a smooth-enough and positive function with an integral equal to one.

- In peak mode, the taylorgram is a sum of Gaussian curves, each curve being associated with a hydrodynamic radius $R_h$:

$$\boldsymbol{S}(\boldsymbol{t}) = \int_0^\infty \frac{S(t_0)}{\sqrt{R_h}} \exp\left(-\frac{2k_B T}{\pi R_c^2 t_0 \eta R_h}\,(t-t_0)^2\right) P(R_h) dR_h$$

- In front mode, the taylorgram is a sum of error functions curves, each curve being associated with a size.

$$\boldsymbol{S}(\boldsymbol{t}) = \int_0^{\infty} \frac{S(t_0)}{2} \left( 1 + erf\left( \sqrt{\frac{2k_B T}{\pi R_c^2 t_0 \eta R_h}} (t - t_0) \right) \right) P(R_h) dR_h$$

In those equations, $R_c$ is the capillary radius, $\eta$ the buffer viscosity, $k_B$ the Boltzmann constant, $T$ the temperature, $t_0$ the elution time and $P(R_h)$ the hydrodynamic probability density function.

In practice, experimental data were sampled at discrete time intervals determined by the detector acquisition rate. This requires the taylorgram to be discretized both in time and in size, with $N_t$ being the number of time steps, and $N_{R_h}$ the number of hydrodynamic radii:

$$s_k = \sum_{j=1}^{N_{R_h}} A_{kj} P_j + \epsilon_k \qquad\qquad k = 1, \dots, N_t$$

with:

$$A_{k,j} = \sum_{j=1}^{N_{R_h}} \frac{c_j}{\sqrt{R_h^j}} \exp\left( -\frac{2k_B T}{\pi R_c^2 t_0 \eta R_h^j} \; (t_k - t_0)^2 \right) \text{ for the peak mode,}$$

$$A_{k,j} = \sum_{j=1}^{N_{R_h}} \frac{c_j}{2} \left( 1 + erf\left( \sqrt{\frac{2k_B T}{\pi R_c^2 t_0 \eta R_h^j}} (t_k - t_0) \right) \right) \text{ for the front mode;}$$

where $c_j$ is a normalizing constant.

The CRLI algorithm then consists into minimizing the following expression:

$$\sum_{k=1}^{N_t} \left( s_k^{exp} - \sum_{j=1}^{N_{R_h}} A_{kj} \, P_j \right)^2 + \alpha \sum_{j=1}^{N_{R_h}-1} (P_{j-1} - 2P_j + P_{j+1})^2$$

where the first term is the squared deviation between the experimental data and the fit, and the second term corresponds to the penalization of unsmoothed probability density functions, seen as a regularization term. $\alpha$ is then called the regularization parameter. Its careful choice allows the regularization term to have an impact on the solution without significantly degrading the fit.

Each taylorgram was analyzed individually and normalized, with the amplitude of the size distribution thus reflecting the relative proportions within that sample, rather than absolute quantities. Across successive measurements, only the mean size and its associated standard deviation were thus be derived and compared. Cipelletti *et al.*[52] showed that this method can differentiate two distinct populations if their sizes differ by at least seven fold the relative width of the population distributions. In our hands, for peak mode, we obtained size distributions with a standard deviation of ~0.5 nm for sizes around 3 nm (relative width ~0.17), meaning the algorithm cannot distinguish populations differing by less than ~1 nm. For front mode, this standard deviation was around 2 nm for sizes around 3 nm (relative width ~0.7), meaning populations cannot be resolved if they differ by less than ~4.5 nm.

Beyond the analysis of elution peaks or front-mode profiles, spikes representing large particles with flow-focused trajectories, were selected and analysed based on: (i) their low sampling data and shape (typically ~3-4 data points at our resolution of ~80 ms) ; and (ii) significant fluorescence intensities above the baseline noise (typically 0.001 RFU) (Fig. S1).

**Thioflavin T fluorescence assays.** Thioflavin T (ThT), an amyloid dye binding to hydrophobic cavities within amyloid structures[24,25], was used to monitor the aggregation of PSMα3 peptides. Fluorescence measurements ($\lambda_{excitation}$ = 449 nm / $\lambda_{emission}$ = 482 nm) were acquired on a CLARIOstar plus (BMG Labtech) plate reader, using 384 well flat-bottom black plates (Greiner Bio-One), covered with a transparent lid to avoid evaporation. Each well of 40 µL contained (i) 200 µmol/L ThT prepared in the appropriate medium from a 10 mmol/L stock solution in MilliQ water, and (ii) 50 µmol/L PSMα3 prepared as described above. Measurements were performed at 25 °C, in technical and biological triplicates, every three minutes with a 300 rpm orbital shaking for 15 seconds before each cycle.

**Transmission Electron Microscopy.** PSMα3 peptides were incubated overnight at 37°C, at a final concentration of 50 µmol/L, in a 10 mmol/L sodium phosphate buffer with 150 mmol/L NaCl (pH 8.0). 4.2 µL of this solution were adsorbed onto glow-discharged carbon coated 300 mesh copper grids for 2 minutes. The excess solution was removed using filter paper. Then, a 1 % uranyl acetate solution was applied to the grids for 30 seconds, blotted with filter paper for negative staining. This step was repeated twice, and the samples were left to dry fro few minutes before observations. A CM120 electron microscope operating at 120 kV with an LaB6 filament was used to examine different areas of the grids, and image a representative overview of the species deposited on the grids.

**Atomic Force Microscopy.** An overnight incubated solution of 50 µmol/L PSMα3 was centrifuged for 20 minutes at 14000 rpm, and the pellet was re-suspended in MilliQ water. 10 µL of the solution was deposited on freshly cleaved mica for 30 minutes, rinsed with MilliQ water and left to dry. For the mutant peptides, that did not aggregate, the overnight incubated sample was directly used without centrifugation. AFM topographic images were obtained on a Dimension Fast-Scan setup (Bruker), using the PeakForce Quantitative Nano-Mechanics (PF-QNM) mode with Nitride-coated silicon cantilevers (SNL-C, Bruker). Cantilevers were calibrated before experiment using the thermal noise method (nominal spring constant of 0.24 N/m, tip radius of 2 nm). Images were analysed and processed with the Nanoscope Analysis software (Bruker).

## Results

### Principle of measurements

In solution, as any micro-/nano-objects, peptides undergo Brownian motion due to thermal agitation, with a diffusion constant:

$$D_0 = {k_B T}/{(6\pi\eta R_h)}$$

where $k_B$ is the Boltzmann constant, $T$ the temperature, $\eta$ the solvent viscosity and $R_h$ the hydrodynamic radius. If this solution is placed in a microfluidic channel under Poiseuille laminar flow, an advection-diffusion coupling further takes place, and leads to an effective dispersion enhancement, known as Taylor dispersion[32,33] (Fig. 1A). In this case, the late-time effective diffusion constant becomes:

$$\mathrm{D} = \mathrm{D_0}(1 + \alpha \mathrm{Pe}^2) \quad \text{with} \quad \mathrm{Pe} = \frac{\mathrm{R_c U}}{\mathrm{D_0}}$$

where Pe is the Peclet number, $R_c$ the channel radius, $U$ the average flow speed, and $\alpha$ a numerical geometric constant. Consequently, the hydrodynamic radius $R_h$ of the peptide can be measured by the simple concentration spreading without the need for any single-particle tracking or fluorescent label.

In practise, a discrete plug of peptides is injected into a capillary (peak mode) and is detected, after a time delay (or elution time), $t_0$, by an ultraviolet (UV) or fluorescence detector placed at a defined position. The combination of the parabolic velocity profile and the molecular diffusion redistributes the peptides perpendicularly to the capillary, resulting in an elution profile called taylorgram (Fig. 1A). For a monodisperse solution, the temporal variance, $\sigma_t^2$, of the taylorgram gives direct access to the diffusion constant, and thus the hydrodynamic size:

$$D = {(R_c^2 t_0)}\Big/{(24\sigma_t^2)}$$

In large Pe regime, while large objects diffuse slowly and are thus more prone to dispersion, leading to broader elution peak, small objects diffuse faster, leading a narrower elution peak.

## Experimental optimisation to detect fluorescence

To establish the feasibility of Phe-based detection in FIDA, we first optimized experimental conditions, focusing on (i) peptide concentration to detect significant fluorescence, (ii) and salt concentration in the mobilizing buffer to minimize peptide stickiness to capillary walls due to the peptide positive charge (Fig. S2). For this purpose, we used a mutant of PSMα3, in which the phenylalanine at position 3 is replaced by an alanine (F3A), rendering it unable to self-assemble into amyloid fibrils[45,46]. Under physiological conditions, in a 20 mmol/L sodium phosphate buffer containing 100 mM NaCl (pH 7.4), no exploitable Gaussian elution profile was obtained; instead, a constant fluorescence plateau was observed after reaching maximal intensity. This is likely due to substantial adsorption to the capillary walls that induces a partial, even complete, loss of symmetry in the Gaussian distribution typically expected. Increasing the ionic strength using 500 mmol/L NaCl reduced the phenomenon, restoring the characteristic Gaussian profile upon peptide injection, though a tailing effect is clearly observed at the end of the taylorgram peak. Despite such mild stickiness, data analysis to quantify the peptide size was performed by adjusting the proportion of the peak that is fitted. Further increasing the salt concentration to 1 mol/L did not improve the signal quality; a salt concentration of 500 mmol/L NaCl was thus selected for the mobilizing to guarantee a good signal quality by reducing significant sticking to the capillary walls.

Next, we injected increasing concentrations of F3A peptides, from 25 µmol/L to 500 µmol/L, to assess the limit of detection for Phe fluorescence, providing that F3A contains four Phe residues (*vs.* five in the WT peptide) (Fig. S3). Concentrations below 100 µM led to weak, if not undetectable, signals with high noise levels, preventing reliable quantification. Conversely, higher concentrations of 250 µmol/L and 500 µmol/L yielded robust, exploitable signals. Under these conditions, we then monitored the F3A peptide in real time for 2 hours at 25°C (Fig. 3, see Fig. S4 for the fits). Despite occasional spikes, and progressive peptide sticking to the capillary in time, the Gaussian profiles remained observable, allowing estimation of the F3A size evolution (Fig. 3A). Like the maximal intensity of the elution profile, the apparent hydrodynamic radius ($R_h$) of F3A remained constant over time after 20 minutes, at ~ 2.3 nm (Fig. 3B, Fig. S5), consistent with the absence of significant aggregation as complementary probed by ThT fluorescence (Fig. 3C) and TEM/AFM observations (Fig. 3D). The unexpected small initial decrease in $R_h$ is likely associated to hydration and/or conformational flexibility of this amphipathic peptide that, upon solvation, might form unstable and tiny clusters before reaching an equilibrium towards an ensemble monomeric state. Besides, as the expected size of the globular protein is ~1.3 nm, this could suggest stable dimerisation upon solubilisation, or more likely a hydrodynamic radius exceeding the theoretical radius for globular protein. To validate that residual sticking did not induce any bias, we also analysed F3A peptides pre-incubated at room temperature for 3 hours (Fig. S6). The hydrodynamic radius of these samples matched that of real-time measurements, confirming the stability of F3A and the absence of any artefact caused by low capillary adhesion. This stable mutant therefore provided an internal calibration for Phe fluorescence, enabling the optimisation of experimental conditions (concentrations, salt, pressure), and quantification of sensitivity and reproducibility in subsequent experiments.

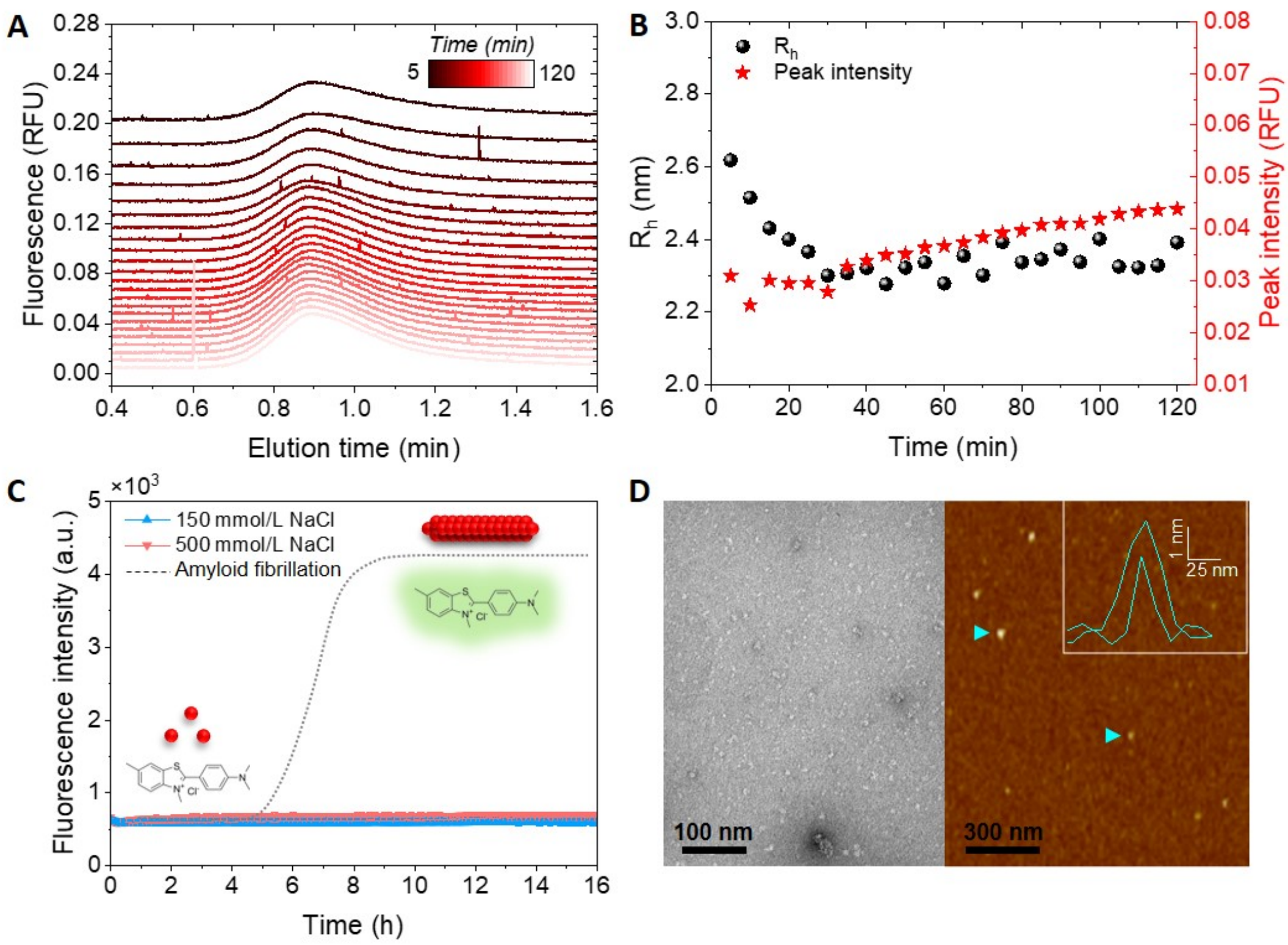

**Figure 3. Peak-mode calibration with a non-amyloid peptide, based on Phe fluorescence. (A)** Elution profiles of the F3A mutant of PSMα3 at different incubation times, from 5 to 120 minutes. Experiments were recorded using (i) a mobile phase of 20 mmol/L sodium phosphate buffer with 500 mmol/L NaCl (pH 7.4), and (ii) a mobilization pressure of 800 mbar. **(B)** Time evolution of the maximal peak intensity and hydrodynamic radius of F3A, derived from the fitting of the taylorgrams. **(C)** Time evolution of Thioflavin T (ThT) fluorescence in presence of F3A in low / high salt conditions (blue/red lines). In dashed grey, a guide for the eye exemplifying a theroretical ThT curve for amyloidogenic proteins, where ThT only binds amyloid structures, thereby emitting fluorescence, as schematically represented. The absence of ThT for F3A indicates no aggregation. **(D)** TEM cliché (left) and AFM topographic image (right) of F3A showing mainly small and amorphous aggregates. Inset. Height profiles of two aggregates.

## Characterization of PSMα3 aggregation in peak mode

Based on the above optimisation, we then investigated the aggregation of the WT PSMα3 peptide, using peak-mode FIDA. The peptide was injected at the optimal concentration of 500 µmol/L (Fig. S7) in a 20 mmol/L sodium phosphate buffer supplemented with 500 mmol/L NaCl (pH 7.4). Measurements were recorded on pre-incubated peptide solutions at defined time points (every 5 to 10 minutes). Aliquots from the same batch, "frozen" at specific time points, were thus injected and analysed in peak mode (Fig. 4, see Fig. S8 for the fits). Regardless of the incubation time, we consistently observed a Gaussian profile – with signs of sticking to the capillary walls (Fig. 3A). The fitting of those profiles unexpectedly revealed a constant hydrodynamic radius of ~2.6 nm over time, yet with a sharpening of the size distribution (Fig. 4B, Fig. S9). Besides, a significant decrease in peak intensity occurred over two hours, with the signal-to-noise ratio preventing any quantification after one hour of aggregation (Fig. 4B). This was consistent with the fibrillation kinetics of PSMα3 probed by ThT fluorescence, with a maximal intensity reached within one to hour, depending on peptide concentration, as well as salt concentration (Fig. 4C, Fig. S10).

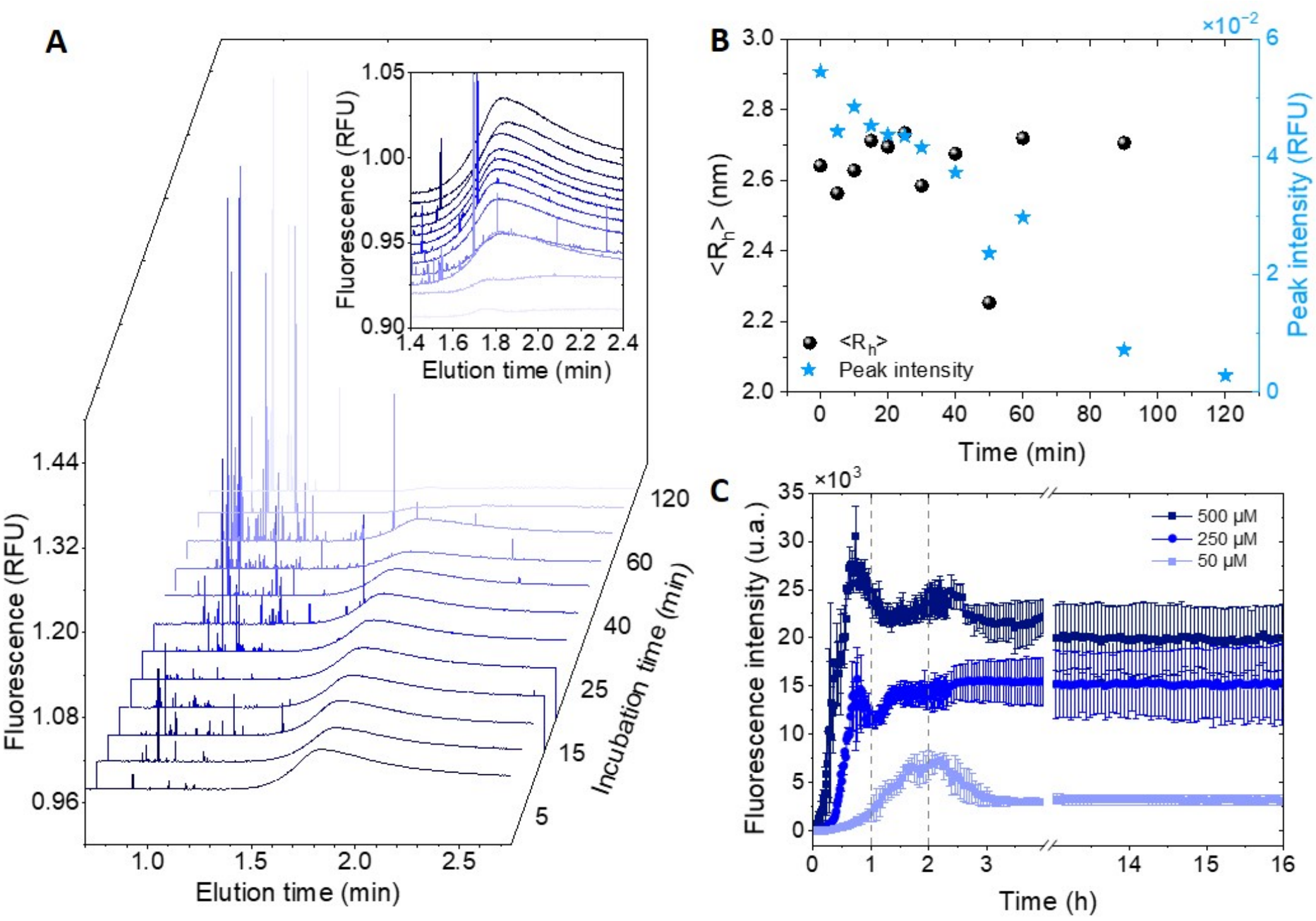

**Figure 4. Fast aggregation of PSMα3 followed by the peak-mode approach. (A)** Elution profiles of the WT-PSMα3 at different incubation times, from 5 to 120 minutes. Experiments were recorded using (i) a mobile phase of 20 mmol/L sodium phosphate buffer with 500 mmol/L NaCl (pH 7.4), and (ii) a mobilization pressure of 400 mbar. **(B)** Time evolution of the maximal peak intensity, correlated with the hydrodynamic radii derived from the Gaussian functions. **(C)** Kinetics of WT, at distinct concentrations, followed by Thioflavin T (ThT) fluorescence.

The drop in fluorescence intensity was accompanied by an increase in the number and/or intensity of the spikes preceding the Gaussian profile in the taylorgram (Fig. 5A). Such spikes, defined by a "Dirac-like" signal, reflect large particles for which radial diffusion becomes negligible resulting in a confinement, and flow-focusing behaviour. Those spikes were likely favoured by the high concentration used (500 µmol/L). However, even at lower concentration of 50 µmol/L, and with limited time evolution after 3 hours, bundles and clusters of thick and long fibrils were mostly observed by both TEM and AFM, suggesting that, whatever the concentration is, the fast-kinetics of PSMα3 aggregation yielded large objects from the early stages (Fig. 5B-C). Altogether, these peak-mode observations thus reflected the depletion of small monomeric/oligomeric species, hereafter named soluble species, as they undergo Taylor dispersion being both radially diffused and advected by the flow leading to a typical elution peak with a Gaussian profile. This depletion is consistent with their conversion into large oligomers and/or fibrils, that get stuck in the Poiseuille flow and cannot explore the capillary through diffusion, adopting a stochastic spike behaviour.

Noteworthy, the static incubation conditions prevented real-time measurements of the same peptide sample as sedimentation occurred within the wells, thus limiting our ability to track the aggregation dynamics.

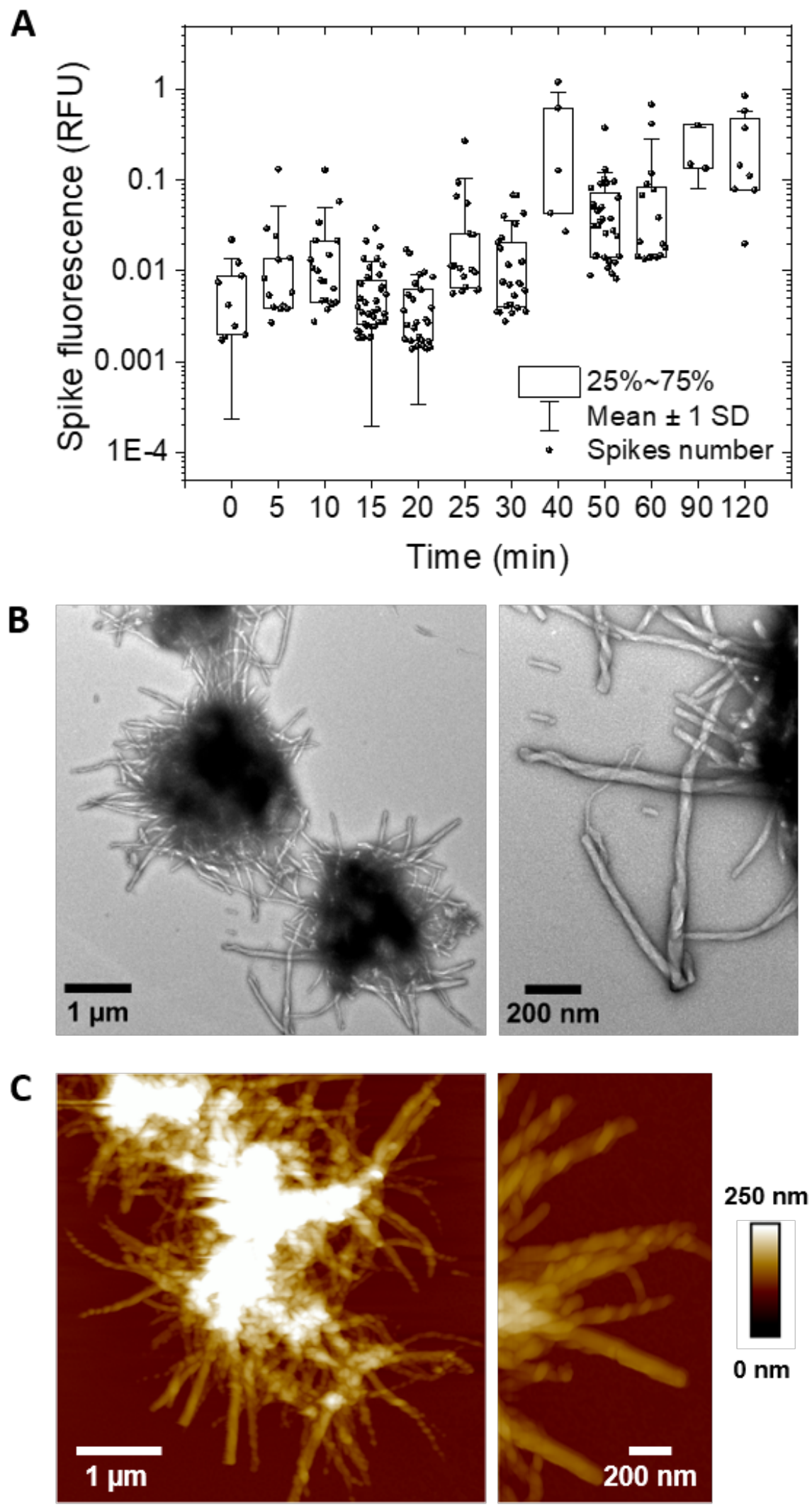

**Figure 5. Characterization of non-diffusive PSMα3 species. (A)** Time evolution of the number and fluorescence intensity of spikes preceding the Gaussian peaks. Same experimental conditions as in Figure 4. **(B)** TEM cliché and **(C)** AFM topographic images of bundles of thick and high PSMα3 ribbon-like and twisted fibrils.

## Development of the front mode for accurate monitoring of aggregation

The high peptide concentrations required to detect Phe fluorescence in peak-mode (500 µmol/L) not only exceed physiological relevance but could also induce distinct - or promote - aggregation pathways, thus complicating the resolution and disentanglement of the various species populations. To address this limitation, we employed an alternative, and so far scarcely used approach, the front mode (elsewhere also named Capflex, for capillary flow experiment). While in peak mode a discrete peptide plug was injected and mobilized in the capillary, in front mode the peptide solution was continuously pushed in the capillary, replacing the existing buffer (Fig. 6A). This approach should thus theoretically allows to reduce the working concentration by increasing the probed volume.

We first performed calibration experiments with the F3A peptide - unable to aggregate - as previously done in peak mode (Fig. 6A). In front mode, instead of a Gaussian profile, we observed a characteristic "front" profile, *i.e.* a sharp increase in fluorescence signal as peptides passed the detector, followed by a plateau at maximal intensity as peptides are pushed through the capillary. Concentrations above 50 µmol/L all yielded significant increase in fluorescence, with a linear increase of the plateau intensity with concentration (Fig. 6B). Since standard amyloid assays on PSMα3 are typically performed at 50 µmol/L, we selected this concentration for subsequent experiments. Importantly, given that F3A peptide does not aggregate, the plateau intensity directly correlates with the concentration of diffusive and soluble species, allowing the establishment of a calibration curve for concentration-dependent fluorescence. F3A peptides exhibited a hydrodynamics radius of ~3.2 nm, at a concentration of 50 µmol/L (Fig. S11), thus larger than the one obtained in peak mode. This discrepancy likely arises from different experimental conditions, namely concentration, mobilizing volume and pressure, that could induce distinct conformational arrangement of the globular monomer.

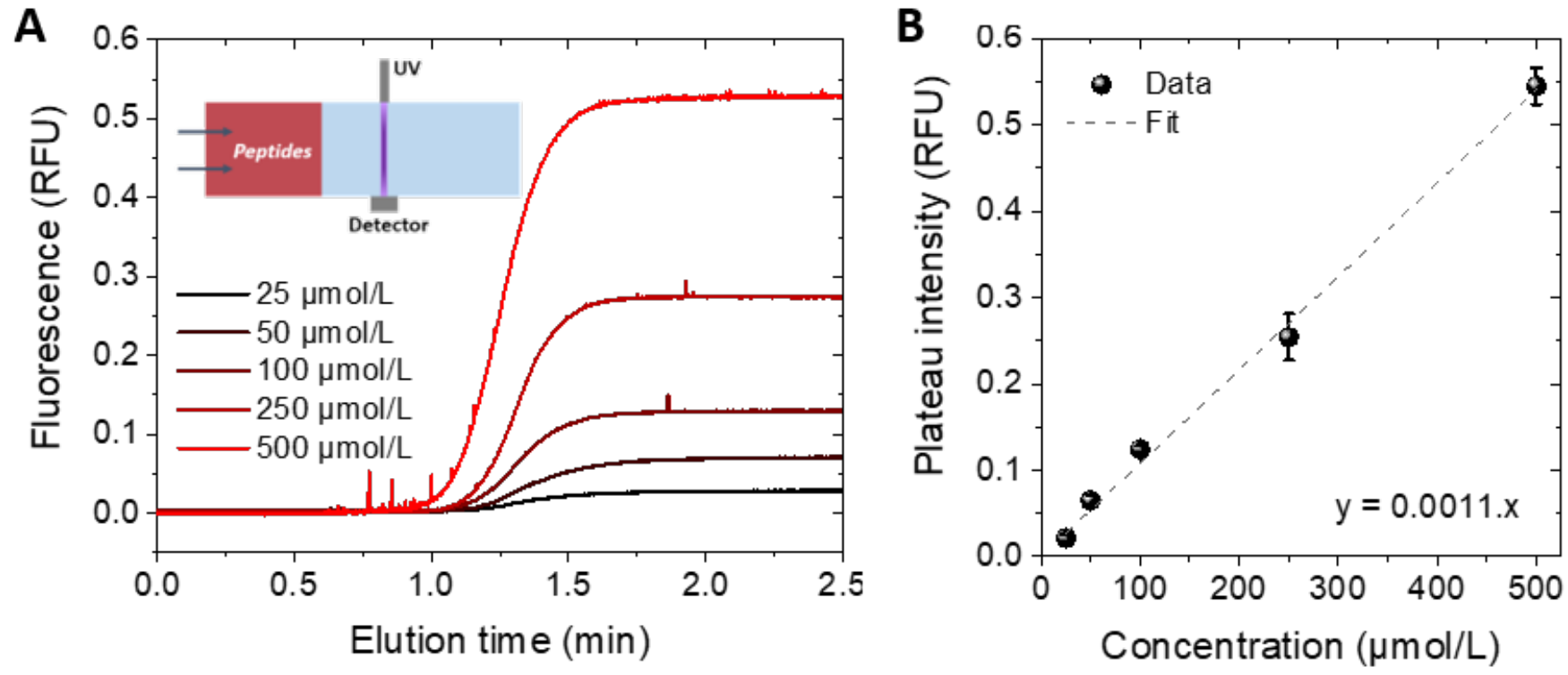

**Figure 6. Front mode to probe soluble species concentration. (A)** Elution profiles, obtained in front mode, of the F3A-PSMα3 at distinct concentrations. Experiments were recorded using (i) a mobile phase of 20 mmol/L sodium phosphate buffer with 500 mmol/L NaCl (pH 7.4), and (ii) a mobilization pressure of 800 mbar. Inset. Schematic representation of the front-mode principle. A concentration step of peptides is injected, resulting in a front profile after a given delay in time. **(B)** Standard curve calibration of the plateau intensity as a function of peptide concentration.

We subsequently monitored pre-incubated solutions of WT PSMα3, at 50 µmol/L, and at defined time points (Fig. S12, Fig. 7). As observed in peak mode, the front elution profiles revealed a time-dependent decrease in the maximal fluorescence intensity - corresponding to the plateau -, indicating the depletion of soluble species probed in the continuously mobilized dilute phase in the capillary (Fig. 7A, see Fig. S13 for the fits). To quantify this behaviour, we applied the F3A calibration curve to WT, revealing a biphasic trend: an initial fast consumption of soluble species within the first hour, followed by a slower rate of consumption during four hours (Fig. 7C). Such consumption was confirmed by Tris-glycine gel electrophoresis experiments, which showed a progressive decrease in intensity of the band around 2-3 kDa, corresponding to mono-/dimeric PSMα3 (Fig. S14). Interestingly, the biphasic depletion behaviour was still observed in real-time measurements, when solutions were sampled from the same well, albeit with distinct consumption rates (Fig. S15). As already suggested by the data in peak mode, these findings underscore the necessity of dynamic conditions, critical for fast-fibrillating systems such as PSMα3, to preserve peptide samples until actual measurement, and prevent aggregation and sedimentation that otherwise bias the species present in the capillary. Markedly, the depletion of soluble species was further corroborated by the progressive appearance of spikes preceding the plateau that, similar to peak-mode observations, reflect insoluble and flow-focused aggregates.

Finally, from these elution profiles, we developed an algorithm approach to infer the size distribution of diffusing species in the capillary. Over time, the average hydrodynamic radius of such species increased from ~2.5 to 4.5 nm, with a broadening of the size distribution. This evolution was consistent with the formation of low molecular weight oligomers, likely present in the extended band in the electrophoretic gel (Fig. 7B, Fig. S14). However, insoluble large oligomers and mature fibrils, captured in the unresolved spikes, remained inaccessible and beyond the detection limits of Taylor dispersion due to their size. To address this, we performed TEM observations on WT PSMα3 at distinct time points (Fig. 7D). They first confirm the progressive disappearance of very small mono- and oligomers but also reveal the existence of both long (> 50 nm) and thin (< 10 nm) filaments and mature thick (> 50nm) fibrils (µm-length) after 30 minutes of incubation.

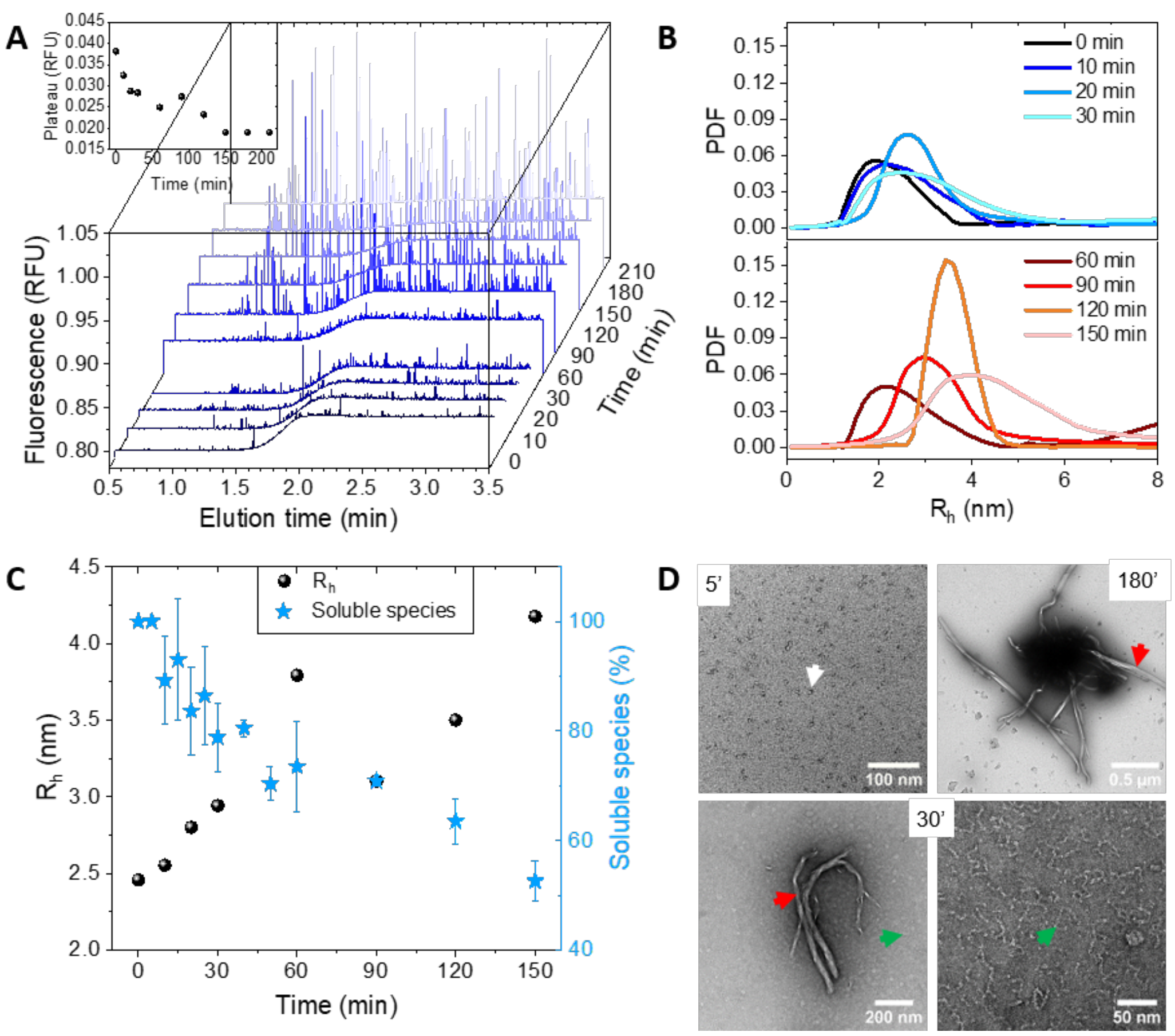

**Figure 7. Front mode captures the fast aggregation of PSMα3. (A)** Elution profiles of the WT-PSMα3 (C = 50 µmol/L) at distinct incubation times. Inset: time evolution of the plateau intensity. Experiments were recorded using (i) a mobile phase of 20 mmol/L sodium phosphate buffer with 500 mmol/L NaCl (pH 7.4), and (ii) a mobilization pressure of 400 mbar. **(B)** Size distribution of the WT at different incubation time, derived from the elution profiles. **(C)** Consumption of WT-PSMα3 soluble species in time, derived from the time evolution of plateau intensities and the standard calibration curve on F3A. **(D)** TEM

images of WT-PSMα3 after 5, 30 and 180 minutes of incubation showing monomers (white arrow), oligomers (green arrow) and mature fibrils (red arrow).

## Discussion & Conclusion

In this work, we establish that the Flow Induced Dispersion Analysis (FIDA) methodology, previously used to characterize canonical amyloid proteins such as Aβ, can be extended to study label-free peptides and proteins lacking tryptophan or tyrosine residues, and prone to fast self-assembly (< 1 hour). To the best of our knowledge, this work constitutes the first demonstration of quantitative analysis of amyloid aggregation based solely on intrinsic phenylalanine fluorescence. This approach thus overcomes the limitations of traditional amyloid assays, such as ThT fluorescence spectroscopy of fluorescence microscopy, which rely on extrinsic labels that may alter the aggregation pathways, or require high quantum-yield aromatic residues.

In particular, a key advancement of this study is the use of the front mode, an underestimated approach in FIDA, that offers a promising analytical window to characterize amyloid proteins compared to the classical peak mode, yet scarcely applied to LLPS[37], solubility[47] and micellization[48]. Indeed, by continuously injecting the peptide solution, this mode allows to reduce the required working concentration to near-physiological ones, thus also minimizing potential artefacts such as capillary sticking and sample aggregation upon measurement. The front mode allowed us to quantify the consumption of soluble species, and determine their hydrodynamic radii with enhanced sensitivity, for PSMα3, a peptide whose aggregation process been only scarcely studied due to its fast fibrillation. Our results revealed a biphasic depletion of soluble species, accompanied by a rapid increase in the population of insoluble species (spikes), consistent with complementary observations from conventional amyloid assays - ThT fluorescence, TEM and AFM. Notably, only small species below 5 to 10 nm could be observed, highlighting the ability of the front mode to probe the early stages of amyloid aggregation, which are often challenging to characterize under non-denaturating conditions with traditional methods such as ThT fluorescence, gel electrophoresis or Size Exclusion Chromatography (SEC) experiments.

Importantly, to demonstrate the relevance and robustness of our approach, we applied the FIDA methodologies to a more prototypical amyloid peptide involved in type 2 diabetes, the human Islet Amyloid Polypeptide (hIAPP) that lacks tryptophan, and contains a minimal amount of aromatic residues (2 phenylalanine and a tyrosine, Fig. 8)[49]. The evolution of the front-mode fluorescence followed the same aggregation-dependent trend, resulting in an almost absence of plateau after 2 hours (Fig. 8B), accompanied by a significant proportion of spikes, consistent with the fibrillation kinetics reported in the literature[50]. Unlike PSMα3, a marked decrease in the hydrodynamic radius of hIAPP was derived (Fig. 8C) likely due to conformational rearrangements from disordered monomers into β-sheet early structures, further highlighting our ability to probe low molecular weight oligomers and initiation of amyloid nucleation.

**A** **hIAPP:** KCNTATCATQRLANFLVHSSNNFGAILSSTNVGSNTY

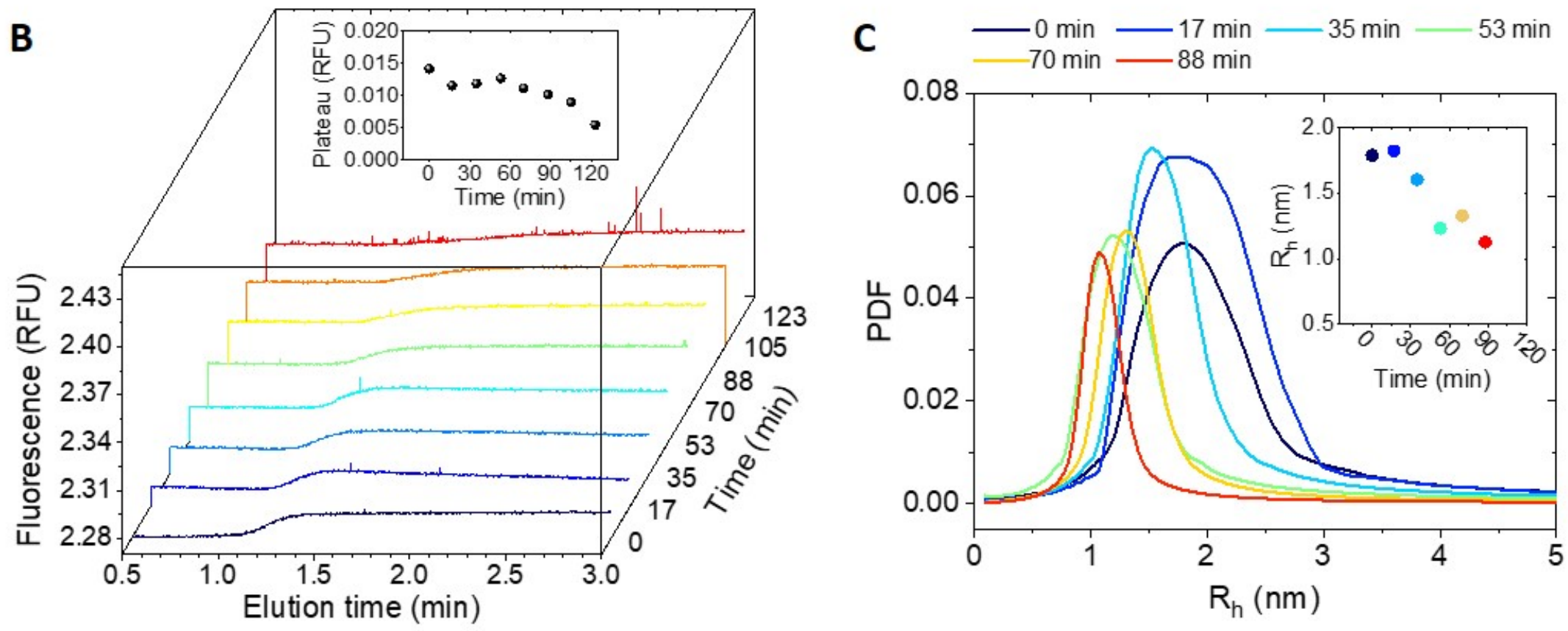

**Figure 8. (A)** Amino-acid sequence of the human Islet Amyloid Polypeptide (hIAPP), where aromatic residues are highlighted in red. **(B)** Elution profiles of hIAPP during real-time front-mode experiments, recorded from the same sample solution (C = 50 µmol.L). Inset: time evolution of the plateau intensity. Experiments were recorded using (i) a mobile phase of 20 mmol/L sodium phosphate buffer with 500 mmol/L NaCl (pH 7.4), and (ii) a mobilization pressure of 800 mbar. **(C)** Size distribution of hIAPP at different representative times. Inset: time evolution of hydrodynamic radii obtained from the size distributions.

The main limitation of our approach lies in the inability to characterize insoluble aggregates, which appear as spikes in the elution profiles. These species, either formed before or after injection, are indeed excluded from the Taylor dispersion regime and cannot be properly analysed. Addressing this challenge would require the development of a dedicated theoretical framework. Yet, when combined with complementary techniques, such as ThT fluorescence, TEM and AFM observations which can probe those large oligomeric and fibrillar structures, the front-mode based FIDA approach provides a comprehensive picture of amyloid speciation. It offers quantitative insights into the proportion of early soluble species relative to mature fibrils. Such an integrated approach is undeniably necessary to enhance our understanding of amyloid formation, and, coupled to structural analysis, could help in identifying potentially toxic intermediates, as early oligomeric species are recognized as therapeutic targets.

## Supporting Information

Additional figures (Figures S1-S15) as mentioned in the text.

## Acknowledgements

The authors first gratefully acknowledge Fidabio for their technical help and advices, and fruitful discussions. They acknowledge financial support from the European Union through the European Research Council under PUMBA (ERC StG 101162069) and EMetBrown (ERC CoG 101039103) grants. Views and opinions expressed are however those of the authors only and do not necessarily reflect those of the EU. Neither the EU nor the granting authority can be held responsible for them. The authors also acknowledge financial support from the Interdisciplinary and Exploratory Research program at University of Bordeaux (MISTIC grant), and the Réseaux de Recherche Impulsion “Frontiers of Life. Finally, they thank the Soft Matter Collaborative Research Unit, Frontier Research Center for Advanced Material and Life Science, Faculty of Advanced Life Science, Hokkaido University, Sapporo, Japan, and the CNRS International Research Network between France and India on "Hydrodynamics at small scales: from soft matter to bioengineering".

## Author Contributions

L.M. performed experimental research on FIDA and G.A. conducted data analysis with algorithms implementation. L.B. performed biophysical characterizations (ThT, TEM and AFM). M.L. helped with data analysis, and algorithm development. M.M.G and T.S. conceived and led the project, secured funding and contributed to data acquisition and analysis. Y.A. and L.K. assisted with data analysis and interpretation. M.M.G and G.A. wrote the manuscript. All authors reviewed, edited, and approved the final version.

**Conflicts of interest.** The authors declare no conflict of interest

## References

(1) Chun Ke, P.; Sani, M.-A.; Ding, F.; Kakinen, A.; Javed, I.; Separovic, F.; P. Davis, T.; Mezzenga, R. Implications of Peptide Assemblies in Amyloid Diseases. *Chemical Society Reviews* **2017**, *46* (21), 6492–6531. https://doi.org/10.1039/C7CS00372B.

(2) Eisenberg, D.; Jucker, M. The Amyloid State of Proteins in Human Diseases. *Cell* **2012**, *148* (6), 1188–1203. https://doi.org/10.1016/j.cell.2012.02.022.

(3) Knowles, T. P. J.; Vendruscolo, M.; Dobson, C. M. The Amyloid State and Its Association with Protein Misfolding Diseases. *Nat Rev Mol Cell Biol* **2014**, *15* (6), 384–396. https://doi.org/10.1038/nrm3810.

(4) Selkoe, D. J. Cell Biology of Protein Misfolding: The Examples of Alzheimer's and Parkinson's Diseases. *Nature Cell Biology* **2004**, *6* (11), 1054–1061. https://doi.org/10.1038/ncb1104-1054.

(5) Fowler, D. M.; Koulov, A. V.; Balch, W. E.; Kelly, J. W. Functional Amyloid – from Bacteria to Humans. *Trends in Biochemical Sciences* **2007**, *32* (5), 217–224. https://doi.org/10.1016/j.tibs.2007.03.003.

(6) Chiti, F.; Dobson, C. M. Protein Misfolding, Functional Amyloid, and Human Disease. *Annual Review of Biochemistry* **2006**, *75* (1), 333–366. https://doi.org/10.1146/annurev.biochem.75.101304.123901.

(7) Van Gerven, N.; Van der Verren, S. E.; Reiter, D. M.; Remaut, H. The Role of Functional Amyloids in Bacterial Virulence. *Journal of Molecular Biology* **2018**, *430* (20), 3657–3684. https://doi.org/10.1016/j.jmb.2018.07.010.

(8) Schwartz, K.; Boles, B. R. Microbial Amyloids – Functions and Interactions within the Host. *Current Opinion in Microbiology* **2013**, *16* (1), 93–99. https://doi.org/10.1016/j.mib.2012.12.001.

(9) Shanmugam, N.; Baker, M. O. D. G.; Ball, S. R.; Steain, M.; Pham, C. L. L.; Sunde, M. Microbial Functional Amyloids Serve Diverse Purposes for Structure, Adhesion and Defence. *Biophys Rev* **2019**, *11* (3), 287–302. https://doi.org/10.1007/s12551-019-00526-1.

(10) Nelson, R.; Sawaya, M. R.; Balbirnie, M.; Madsen, A. Ø.; Riekel, C.; Grothe, R.; Eisenberg, D. Structure of the Cross-β Spine of Amyloid-like Fibrils. *Nature* **2005**, *435* (7043), 773–778. https://doi.org/10.1038/nature03680.

(11) Jahn, T. R.; Makin, O. S.; Morris, K. L.; Marshall, K. E.; Tian, P.; Sikorski, P.; Serpell, L. C. The Common Architecture of Cross-β Amyloid.

(12) Karran, E.; Mercken, M.; Strooper, B. D. The Amyloid Cascade Hypothesis for Alzheimer's Disease: An Appraisal for the Development of Therapeutics. *Nat Rev Drug Discov* **2011**, *10* (9), 698–712. https://doi.org/10.1038/nrd3505.

(13) Ricciarelli, R.; Fedele, E. The Amyloid Cascade Hypothesis in Alzheimer's Disease: It's Time to Change Our Mind. *Current Neuropharmacology* **2017**, *15* (6), 926–935. https://doi.org/10.2174/1570159X15666170116143743.

(14) Siddiqi, M. K.; Malik, S.; Majid, N.; Alam, P.; Khan, R. H. Chapter Ten - Cytotoxic Species in Amyloid-Associated Diseases: Oligomers or Mature Fibrils. In *Advances in Protein Chemistry and Structural Biology*; Donev, R., Ed.; Protein Misfolding; Academic Press, 2019; Vol. 118, pp 333–369. https://doi.org/10.1016/bs.apcsb.2019.06.001.

(15) Abedini, A.; Plesner, A.; Cao, P.; Ridgway, Z.; Zhang, J.; Tu, L.-H.; Middleton, C. T.; Chao, B.; Sartori, D. J.; Meng, F.; Wang, H.; Wong, A. G.; Zanni, M. T.; Verchere, C. B.; Raleigh, D. P.; Schmidt, A. M. Time-Resolved Studies Define the Nature of Toxic IAPP Intermediates, Providing Insight for Anti-Amyloidosis Therapeutics. *eLife* **2016**, *5*, e12977. https://doi.org/10.7554/eLife.12977.

(16) Zhaliazka, K.; Ali, A.; Kurouski, D. Phospholipids and Cholesterol Determine Molecular Mechanisms of Cytotoxicity of α-Synuclein Oligomers and Fibrils. *ACS Chem. Neurosci.* **2024**. https://doi.org/10.1021/acschemneuro.3c00671.

(17) De, S.; Wirthensohn, D. C.; Flagmeier, P.; Hughes, C.; Aprile, F. A.; Ruggeri, F. S.; Whiten, D. R.; Emin, D.; Xia, Z.; Varela, J. A.; Sormanni, P.; Kundel, F.; Knowles, T. P. J.; Dobson, C. M.; Bryant, C.; Vendruscolo, M.; Klenerman, D. Different Soluble Aggregates of Aβ42 Can Give Rise to Cellular Toxicity through Different Mechanisms. *Nat Commun* **2019**, *10*, 1541. https://doi.org/10.1038/s41467-019-09477-3.

(18) Flach, K.; Hilbrich, I.; Schiffmann, A.; Gärtner, U.; Krüger, M.; Leonhardt, M.; Waschipky, H.; Wick, L.; Arendt, T.; Holzer, M. Tau Oligomers Impair Artificial Membrane Integrity and Cellular Viability*. *Journal of Biological Chemistry* **2012**, *287* (52), 43223–43233. https://doi.org/10.1074/jbc.M112.396176.

(19) Lenz, M.; Witten, T. A. Geometrical Frustration Yields Fibre Formation in Self-Assembly. *Nature Phys* **2017**, *13* (11), 1100–1104. https://doi.org/10.1038/nphys4184.

(20) Ragonis-Bachar, P.; Landau, M. Functional and Pathological Amyloid Structures in the Eyes of 2020 Cryo-EM. *Current Opinion in Structural Biology* **2021**, *68*, 184–193. https://doi.org/10.1016/j.sbi.2021.01.006.

(21) Loquet, A.; El Mammeri, N.; Stanek, J.; Berbon, M.; Bardiaux, B.; Pintacuda, G.; Habenstein, B. 3D Structure Determination of Amyloid Fibrils Using Solid-State NMR Spectroscopy. *Methods* **2018**, *138–139*, 26–38. https://doi.org/10.1016/j.ymeth.2018.03.014.

(22) Adamcik, J.; Mezzenga, R. Study of Amyloid Fibrils via Atomic Force Microscopy. *Current Opinion in Colloid & Interface Science* **2012**, *17* (6), 369–376. https://doi.org/10.1016/j.cocis.2012.08.001.

(23) Rodriguez, A.; Ali, A.; Holman, A. P.; Dou, T.; Zhaliazka, K.; Kurouski, D. Nanoscale Structural Characterization of Transthyretin Aggregates Formed at Different Time Points of Protein Aggregation Using Atomic Force Microscopy-infrared Spectroscopy. *Protein Sci* **2023**, *32* (12), e4838. https://doi.org/10.1002/pro.4838.

(24) Xue, C.; Lin, T. Y.; Chang, D.; Guo, Z. Thioflavin T as an Amyloid Dye: Fibril Quantification, Optimal Concentration and Effect on Aggregation. *R. Soc. open sci.* **2017**, *4* (1), 160696. https://doi.org/10.1098/rsos.160696.

(25) Groenning, M. Binding Mode of Thioflavin T and Other Molecular Probes in the Context of Amyloid Fibrils—Current Status. *J Chem Biol* **2010**, *3* (1), 1–18. https://doi.org/10.1007/s12154-009-0027-5.

(26) Rovnyagina, N. R.; Budylin, G. S.; Vainer, Y. G.; Tikhonova, T. N.; Vasin, S. L.; Yakovlev, A. A.; Kompanets, V. O.; Chekalin, S. V.; Priezzhev, A. V.; Shirshin, E. A. Fluorescence Lifetime and Intensity of Thioflavin T as Reporters of Different Fibrillation Stages: Insights Obtained from Fluorescence Up-Conversion and Particle Size Distribution Measurements. *IJMS* **2020**, *21* (17), 6169. https://doi.org/10.3390/ijms21176169.

(27) Castello, F.; Paredes, J. M.; Ruedas-Rama, M. J.; Martin, M.; Roldan, M.; Casares, S.; Orte, A. Two-Step Amyloid Aggregation: Sequential Lag Phase Intermediates. *Sci Rep* **2017**, *7*, 40065. https://doi.org/10.1038/srep40065.
(28) Rice, L. J.; Ecroyd, H.; van Oijen, A. M. Illuminating Amyloid Fibrils: Fluorescence-Based Single-Molecule Approaches. *Comput Struct Biotechnol J* **2021**, *19*, 4711–4724. https://doi.org/10.1016/j.csbj.2021.08.017.
(29) Teale, F. W. J.; Weber, G. Ultraviolet Fluorescence of the Aromatic Amino Acids. *Biochem J* **1957**, *65* (3), 476–482. https://doi.org/10.1042/bj0650476.
(30) Ghisaidoobe, A. B. T.; Chung, S. J. Intrinsic Tryptophan Fluorescence in the Detection and Analysis of Proteins: A Focus on Förster Resonance Energy Transfer Techniques. *International Journal of Molecular Sciences* **2014**, *15* (12), 22518–22538. https://doi.org/10.3390/ijms151222518.
(31) Pedersen, M. E.; Østergaard, J.; Jensen, H. Flow-Induced Dispersion Analysis (FIDA) for Protein Quantification and Characterization. In *Clinical Applications of Capillary Electrophoresis: Methods and Protocols*; Phillips, T. M., Ed.; Springer: New York, NY, 2019; pp 109–123. https://doi.org/10.1007/978-1-4939-9213-3_8.
(32) Aris, R. On the Dispersion of a Solute in a Fluid Flowing through a Tube. *Proc. A* **1956**, *235* (1200), 67–77. https://doi.org/10.1098/rspa.1956.0065.
(33) Taylor, G. I. Dispersion of Soluble Matter in Solvent Flowing Slowly through a Tube. *Proc. A* **1953**, *219* (1137), 186–203. https://doi.org/10.1098/rspa.1953.0139.
(34) Moser, M. R.; Baker, C. A. Taylor Dispersion Analysis in Fused Silica Capillaries: A Tutorial Review. *Anal. Methods* **2021**, *13* (21), 2357–2373. https://doi.org/10.1039/D1AY00588J.
(35) Deleanu, M.; Hernandez, J.-F.; Cipelletti, L.; Biron, J.-P.; Rossi, E.; Taverna, M.; Cottet, H.; Chamieh, J. Unraveling the Speciation of β-Amyloid Peptides during the Aggregation Process by Taylor Dispersion Analysis. *Anal. Chem.* **2021**, *93* (16), 6523–6533. https://doi.org/10.1021/acs.analchem.1c00527.
(36) Deleanu, M.; Deschaume, O.; Cipelletti, L.; Hernandez, J.-F.; Bartic, C.; Cottet, H.; Chamieh, J. Taylor Dispersion Analysis and Atomic Force Microscopy Provide a Quantitative Insight into the Aggregation Kinetics of Aβ (1–40)/Aβ (1–42) Amyloid Peptide Mixtures. *ACS Chem. Neurosci.* **2022**, *13* (6), 786–795. https://doi.org/10.1021/acschemneuro.1c00784.
(37) Farzadfard, A.; Kunka, A.; Mason, T. O.; Larsen, J. A.; Norrild, R. K.; Dominguez, E. T.; Ray, S.; Buell, A. K. Thermodynamic Characterization of Amyloid Polymorphism by Microfluidic Transient Incomplete Separation. *Chem. Sci.* **2024**, *15* (7), 2528–2544. https://doi.org/10.1039/D3SC05371G.
(38) Stender, E. G. P.; Ray, S.; Norrild, R. K.; Larsen, J. A.; Petersen, D.; Farzadfard, A.; Galvagnion, C.; Jensen, H.; Buell, A. K. Capillary Flow Experiments for Thermodynamic and Kinetic Characterization of Protein Liquid-Liquid Phase Separation. *Nat Commun* **2021**, *12* (1), 7289. https://doi.org/10.1038/s41467-021-27433-y.
(39) Aragonès Pedrola, J.; Dekker, F. A.; Garfagnini, T.; Mayer, G.; Koopman, M. B.; Bergmeijer, M.; Heesink, G.; Rots, I.; Claessens, M. M. A. E.; Förster, F.; Hoozemans, J. J. M.; Jensen, H.; Friedler, A.; Rüdiger, S. G. D. FibrilPaint to Determine the Length of Tau Amyloids in Fluids. *Proceedings of the National Academy of Sciences* **2025**, *122* (44), e2502847122. https://doi.org/10.1073/pnas.2502847122.
(40) Deleanu, M.; Hernandez, J.-F.; Cottet, H.; Chamieh, J. Taylor Dispersion Analysis Discloses the Impairment of Aβ Peptide Aggregation by the Presence of a Fluorescent Tag. *ELECTROPHORESIS* **2023**, *44* (7–8), 701–710. https://doi.org/10.1002/elps.202200192.
(41) Otto, M. Phenol-Soluble Modulins. *Int. J. Med. Microbiol.* **2014**, *304* (2), 164–169. https://doi.org/10.1016/j.ijmm.2013.11.019.
(42) Peschel, A.; Otto, M. Phenol-Soluble Modulins and Staphylococcal Infection. *Nature Reviews Microbiology* **2013**, *11* (10), 667–673. https://doi.org/10.1038/nrmicro3110.
(43) Li, S.; Huang, H.; Rao, X.; Chen, W.; Wang, Z.; Hu, X. Phenol-Soluble Modulins: Novel Virulence-Associated Peptides of Staphylococci. *Future Microbiology* **2014**, *9* (2), 203–216. https://doi.org/10.2217/fmb.13.153.

(44) Zaman, M.; Andreasen, M. Modulating Kinetics of the Amyloid-Like Aggregation of S. Aureus Phenol-Soluble Modulins by Changes in pH. *Microorganisms* **2021**, *9* (1), 117. https://doi.org/10.3390/microorganisms9010117.

(45) Tayeb-Fligelman, E.; Tabachnikov, O.; Moshe, A.; Goldshmidt-Tran, O.; Sawaya, M. R.; Coquelle, N.; Colletier, J.-P.; Landau, M. The Cytotoxic Staphylococcus Aureus PSMα3 Reveals a Cross-α Amyloid-like Fibril. *Science* **2017**, *355* (6327), 831–833. https://doi.org/10.1126/science.aaf4901.

(46) Tayeb-Fligelman, E.; Salinas, N.; Tabachnikov, O.; Landau, M. Staphylococcus Aureus PSMα3 Cross-α Fibril Polymorphism and Determinants of Cytotoxicity. *Structure* **2020**, *28* (3), 301-313.e6. https://doi.org/10.1016/j.str.2019.12.006.

(47) Taylor, A. I. P.; Xu, Y.; Wilkinson, M.; Chakraborty, P.; Brinkworth, A.; Willis, L. F.; Zhuravleva, A.; Ranson, N. A.; Foster, R.; Radford, S. E. Kinetic Steering of Amyloid Formation and Polymorphism by Canagliflozin, a Type-2 Diabetes Drug. *J. Am. Chem. Soc.* **2025**, *147* (14), 11859–11878. https://doi.org/10.1021/jacs.4c16743.

(48) Mohammad-Beigi, H.; Mason, T. O.; Rovers, T. A. M.; Jæger, T. C.; Møller, M. S.; Ipsen, R.; Hougaard, A. B.; Svensson, B.; Buell, A. K. Taylor Dispersion Analysis of Micellization (TDAM) Reveals Distinct Assembly and Dissociation Pathways of α-, β-, and κ-Casein Micelles. *Food Hydrocolloids* **2026**, *173*, 112301. https://doi.org/10.1016/j.foodhyd.2025.112301.

(49) Caillon, L.; Hoffmann, A. R. F.; Botz, A.; Khemtemourian, L. Molecular Structure, Membrane Interactions, and Toxicity of the Islet Amyloid Polypeptide in Type 2 Diabetes Mellitus. *Journal of Diabetes Research* **2016**, *2016* (1), 5639875. https://doi.org/10.1155/2016/5639875.

(50) Hoffmann, A. R. F.; Caillon, L.; Vazquez, L. S. S.; Spath, P.-A.; Carlier, L.; Khemtémourian, L.; Lequin, O. Time Dependence of NMR Observables Reveals Salient Differences in the Accumulation of Early Aggregated Species between Human Islet Amyloid Polypeptide and Amyloid-β. *Phys. Chem. Chem. Phys.* **2018**, *20* (14), 9561–9573. https://doi.org/10.1039/C7CP07516B.

(51) Cottet, H.; Biron, J.-P.; Martin, M. On the Optimization of Operating Conditions for Taylor Dispersion Analysis of Mixtures. *Analyst* **2014**, *139* (14), 3552–3562. https://doi.org/10.1039/C4AN00192C.

(52) Cipelletti, L.; Biron, J.-P.; Martin, M.; Cottet, H. Measuring Arbitrary Diffusion Coefficient Distributions of Nano-Objects by Taylor Dispersion Analysis. *Anal. Chem.* **2015**, *87* (16), 8489–8496. https://doi.org/10.1021/acs.analchem.5b02053.

(53) Provencher, S. W. A Constrained Regularization Method for Inverting Data Represented by Linear Algebraic or Integral Equations. *Computer Physics Communications* **1982**, *27* (3), 213–227. https://doi.org/10.1016/0010-4655(82)90173-4.

(54) Provencher, S. W. CONTIN: A General Purpose Constrained Regularization Program for Inverting Noisy Linear Algebraic and Integral Equations. *Computer Physics Communications* **1982**, *27* (3), 229–242. https://doi.org/10.1016/0010-4655(82)90174-6.